# Periodicity of chaotic solutions

M. Berezowski, B. Kulik

Silesian University of Technology, Institute of Mathematics, Gliwice, Poland

E-mail: marek.berezowski@polsl.pl

## Abstract

The scope of the paper is the analysis of the impact of flow reversal on the dynamics of cascades of reactors (Fig.1). Periodic and chaotic oscillations occur in the analyzed system. There is a dependence between the oscillation period of the state variable of the system without flow reversal and the recurrence period of windows of chaos in the steady-state diagram of the system with flow reversal.

## 1. Introduction

The dynamics of reactors was widely discussed in [1-16], where it was proved that in the steady state the concentration and temperature of the flux may oscillate in a manner that is more or less complex. The oscillations may have a multiperiodic, quasi-periodic or chaotic nature, depending on the values of reactor parameters and the concentration and temperature of the reacting flux at the initial state.

The dynamic behaviour of periodically forced chemical reactors (e.g. with cyclically permuting feed and discharge positions or changing the feed flow direction) was studied in [1,4,6-8,11-16]. It was shown that in these networks the type of oscillations also depends on the switching time (i.e. the time interval occurring between two successive permutations of the reactor order or flow reversals, respectively).

This paper is focused on the analysis of the dynamics of a system consisting of two cascades, each containing two tank reactors. It was assumed that the cascades are coupled due to the heat exchange between the reactors (Fig.1). Each cascade is fed with independent flux of raw materials with cyclically reversing the flow directions. In the generalized case, the cycles of the changes are different for both cascades and mutually independent. The calculations results indicated that different types of temperature and concentration oscillations may occur in the analyzed system, including chaotic oscillations. For graphic representation of the results the so-called steady-state diagram was used, following the assumption that the switching time between flow reversals is a bifurcation parameter. Accordingly, in some cases the windows of chaos presented in the diagram appear in a regular manner, i.e. at equal intervals. The intervals were found to correspond exactly to the oscillation period of the state variables of the system without flow reversal. Thus, specific periodicity of the occurrence of the windows of chaos was observed. This phenomenon was previously pointed out in [16], where a single cascade of two reactors with flow reversal was analysed. However, in this paper, the windows of chaos in the bifurcation diagram are proved to occur in a multiperiodic manner, depending on the nature of the time series of the state variables of the system without flow reversal. Accordingly, if the variables oscillate in $N$-periodic manner, the windows of chaos in the steady-state diagram of the system with flow reversal also appear in $N$-periodic manner. Therefore, it is possible to presume that if the state variables of the system without flow reversal oscillated in a non-periodic (chaotic) manner, the windows of chaos in the steady-state diagram of the system with flow reversal would also appear in a non-periodic manner (chaotic).

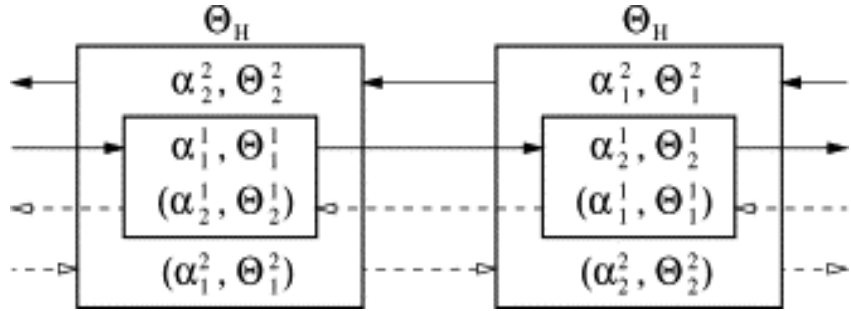

Fig. 1. Scheme of the system with flow reversal. Example of counter-current flow of feed streams.

## 2. Balance model of a dual cascade of two coupled tank reactors with flow reversal

The mathematical model of a dual cascade of chemical tank reactors was subjected to a numerical analysis. It was assumed that each cascade is fed with a separate flux of raw materials with flow reversal (Fig.1). Heat exchange occurs between the cascades, however, the external cascade is cooled by a medium of constant temperature $\Theta_H$. Assuming that each cascade consists of two identical tank reactors, wherein $A \rightarrow B$ chemical reaction of the *n*-th order takes place, the corresponding balance equations for any reactor in such system may be expressed by means of the following balance equations:

Mass balance equation:
$$\frac{d\alpha_{K_x}^x}{d\tau} + \alpha_{K_x}^x - (K_x - 1)\alpha_{K_x-1}^x = \phi_{K_x}^x \quad (1)$$

Heat balance equation
$$\frac{d\Theta_{K_x}^x}{d\tau} + \Theta_{K_x}^x - (K_x - 1)\Theta_{K_x-1}^x = \beta_x \phi_{K_x}^x + (x-1)\delta_H\left(\Theta_H - \Theta_{K_x}^x\right) + (3-2x)\delta\left(\Theta_{K_2}^2 - \Theta_{K_1}^1\right) \quad (2)$$

where *x=1* denotes the internal cascade, whereas *x=2* the external one.

Considering the direction of the flux, constant $K_x$ designates the position of the reactor in the *x-th* cascade at any time. The value of this constant can be calculated from:

$$K_x = K_{x0} + (-1)^{K_{x0}+1}\left[\text{int}\left(\frac{\tau}{\tau_{px}}\right) - 2\,\text{int}\left(\frac{\tau}{2\tau_{px}}\right)\right] \quad (3)$$

where $K_{x0}$ (equal to 1 or 2) denotes the position of a given reactor in the *x-th* cascade at time $\tau = 0$. The co-current flow of both fluxes at given time may be symbolically expressed as: (1;2), (1;2) or (2;1), (2;1), whereas the counter-current flow as: (1;2), (2;1) or (2;1), (1;2). In equation (3), int(*y*) is the integral part of number *y*, whereas $\tau_{px}$ is the time interval occurring between two successive flow reversals in the *x-th* cascade. The function of the reaction kinetics expressed in terms of the balance equations assumes the following form:

$$\phi_{K_x}^x = Da_x \left(1-\alpha_{K_x}^x\right)^{n_x} \exp\left(\gamma_x \frac{\Theta_{K_x}^x}{1+\Theta_{K_x}^x}\right) \qquad (4)$$

## 3. Calculations and results analysis

It was assumed in exemplary calculations that the time intervals occurring between two successive flow reversals are the same for both cascades: $\tau_{p1} = \tau_{p2} = \tau_p$. Furthermore, it was assumed that at the initial stage, the reacting fluxes in both cascades are counter-current. In the curse of the numerical calculations various types of dynamic solutions of the balance model were derived (1)-(4), depending on the values of the reactors parameters and the values of time $\tau_p$. Thus, the obtained solutions, as far as oscillations are concerned, were periodic, multiperiodic and chaotic. The results were presented graphically by means of the steady-state diagrams (Feigenbaum's diagrams), marking on the temperature graph the output flux in the external cascade $\Theta_{2p}^2$ at the times of flow reversals. The assumed bifurcation parameter was the switching time $\tau_p$. Regular repeatability of the occurrence of the windows of chaos as well as periodic oscillations was established. It was proved that this phenomenon is strictly connected with the characteristics of the dynamics of the system without flow reversal, i.e. for $\tau_p = \infty$. Assuming the same values of the parameters for both cascades: $n=1.5$, $\gamma=20$, $\beta=0.95$ and $\delta=3$, $\delta_H=3$, $\Theta_H=-0.02$ and $\tau_p = \infty$ (absence of flow reversal), the steady-state diagram of the counter-currently operating cascades was designated, see Fig.2. The solid lines denote the extremes of the stable steady states, the dashed lines – unstable ones. As seen in the diagram, for *Da=0.06* the system generates stable single-periodic oscillations, see the time series in Fig.3. The oscillation period is *T=14.9*. Upon introducing the flow reversal to the system $(\tau_p < \infty)$, the steady-state diagram was re-designated, this time as a function of switching time $\tau_p$. Consequently, the graph presented in Fig.4. was derived, where the periodically recurrent areas of chaos are clearly discerned, their period is equal to the oscillation period of the state variable depicted in Fig. 3, i.e. *T=14.9*.

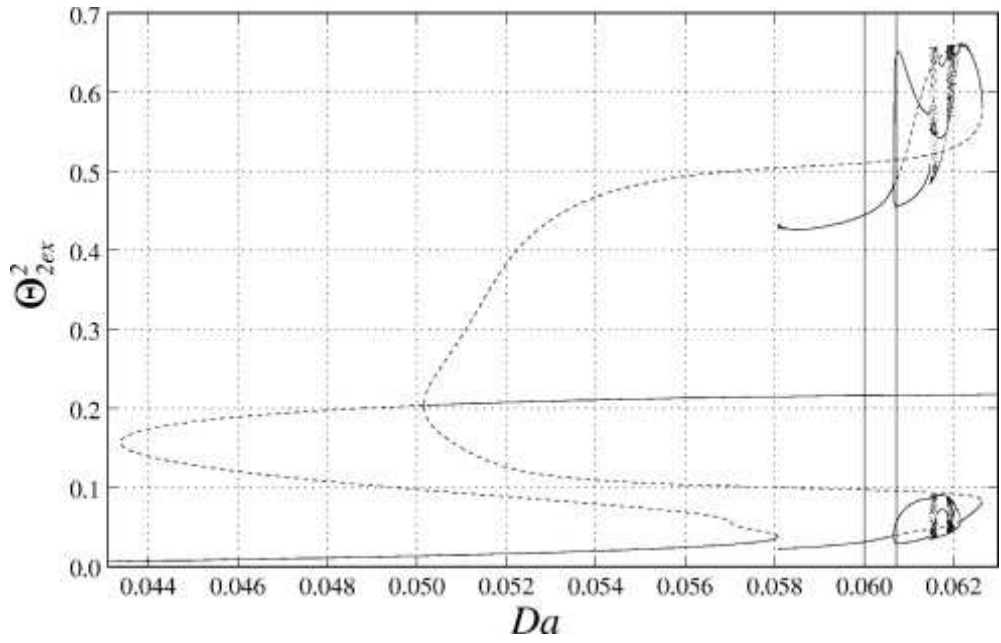

Fig. 2. Steady-state diagram for the system without flow reversal ($\tau_p = \infty$).

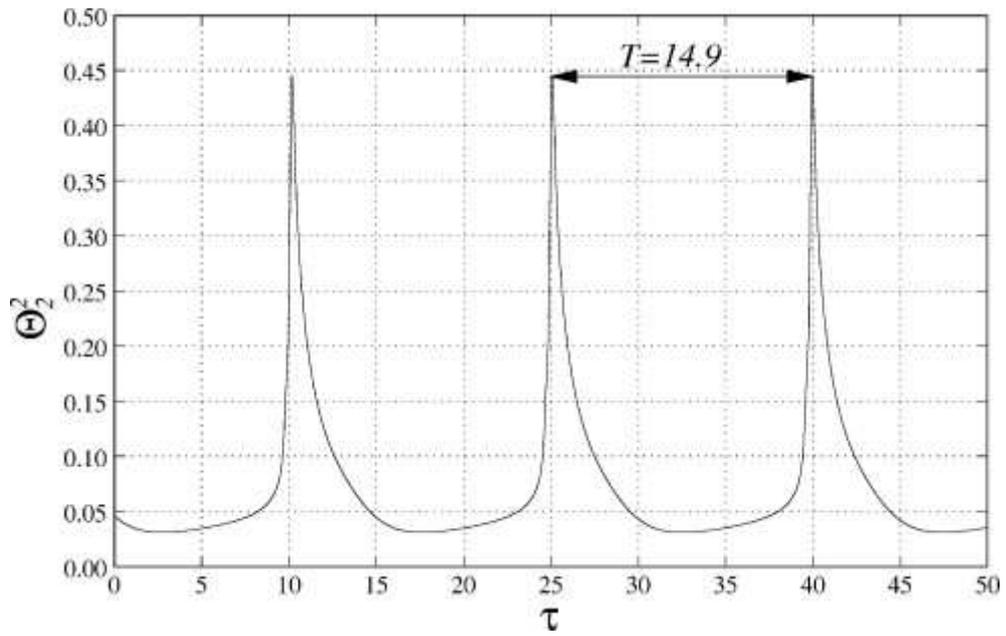

Fig. 3. Time series of the system without flow reversal ($Da = 0.06$, $\tau_p = \infty$).

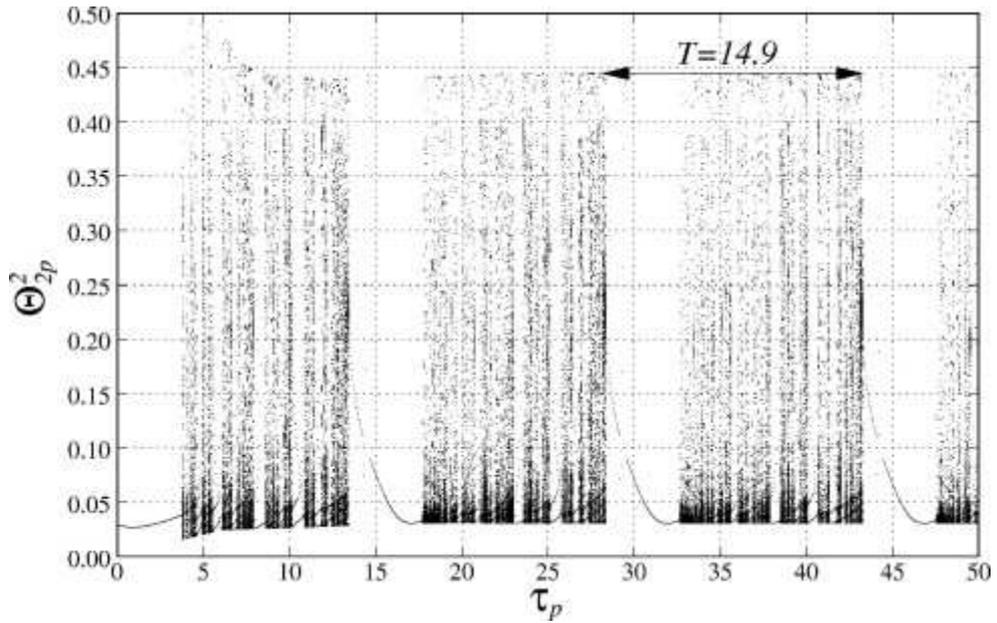

Fig. 4. Steady-state diagram of the system with flow reversal ($Da = 0.06$).

As observable in Fig.2, for $Da=0.0607$ the system without flow reversal ($\tau_p = \infty$) generates stable, two-periodic oscillations, their time series is presented in Fig. 5. The absolute oscillation period is $T=24.1$, whereas the time intervals between the successive maximums are $T_1 =13.4$ and $T_2 =10.7$, respectively. As in the previously discussed case, upon the introduction of the flow reversal ($\tau_p < \infty$) the steady-state diagram was re-designated, as a function of switching time. Consequently, the graph shown in Fig.6. was derived, where two-periodically repeatable areas of chaos are clearly discerned and the distances between the areas of chaos are equal to the above mentioned intervals $T_1 =13.4$, $T_2 =10.7$.

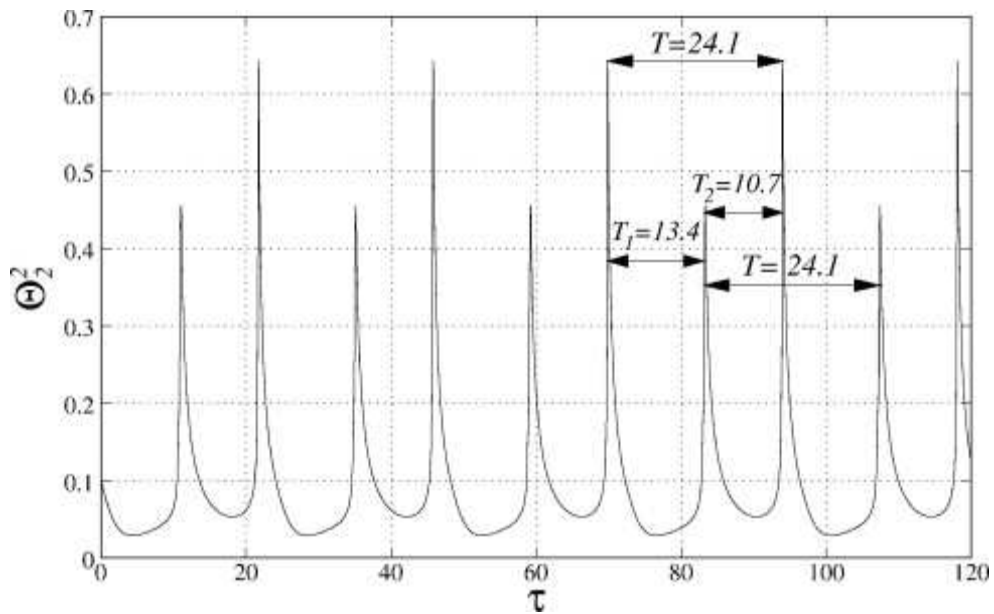

Fig. 5. Time series of the system without flow reversal ($Da = 0.0607$, $\tau_p = \infty$).

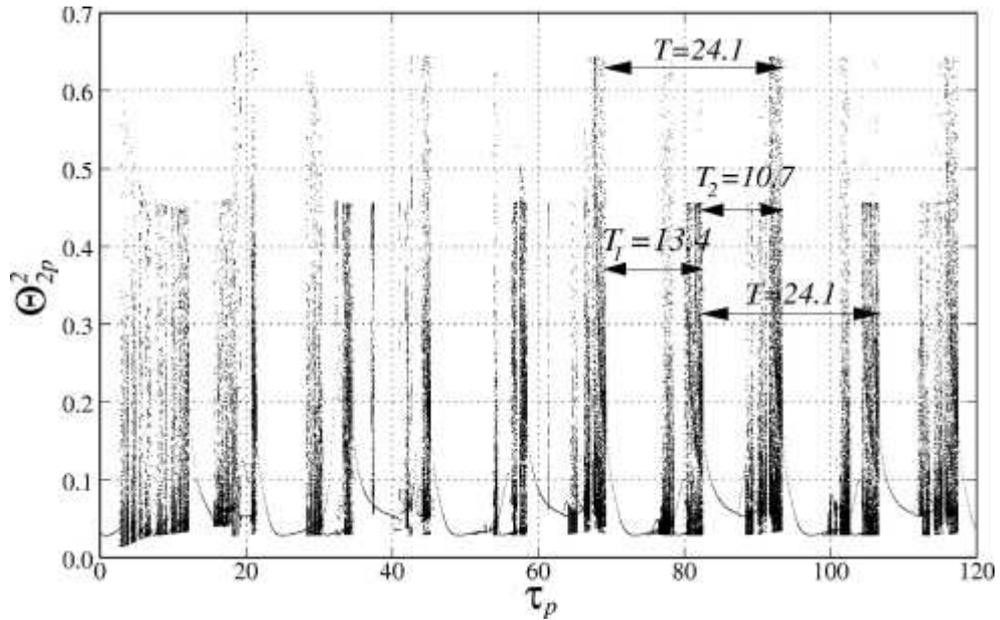

Fig. 6. Steady-state diagram of the system with flow reversal ($Da = 0.0607$).

The explanation of this phenomenon is the following: as seen in Fig.4 and 6, the repeatability of the windows of chaos occurs only for appropriately high values of $\tau_p$. As already mentioned above, if $\tau_p = \infty$ the analysed system generates stable periodic oscillations (single-periodic for $Da=0.06$ and two-periodic for $Da=0.0607$). This means that for appropriately high values of $\tau_p = \tau_p^*$ the time changes in temperature and concentrations will practically stabilize before the next flow reversal. Accordingly, if chaos occurs for $\tau_p = \tau_p^*$, it also occurs for $\tau_p = \tau_p^* + mT$, where $T$ is the oscillation period of the state variables of the system without flow reversal, and $m$ is any natural number. For single-periodic oscillations, identical windows of chaos appear in the steady-state diagram at equal time intervals $\tau_p = \tau_p^* + mT$. For two-periodic oscillations two different chaotic solutions are generated, depending on the phase in the course of which the flow reversal is effected. In consequence, two different windows of chaos are observable in the diagram, situated at the distance equal to the distance between the two phases. As in the previous case, the windows appear at equal time intervals $\tau_p = \tau_p^* + mT$. For $N$-th –periodic time series, the steady-state diagram may indicate $N$ different windows of chaos, repeated at equal time $\tau_p = \tau_p^* + mT$. If $N \to \infty$, infinite number of different windows of chaos may appear in the bifurcation diagram.

## 4. Concluding remarks

The scope of the paper was a numerical analysis of the mathematical model of a double cascade of coupled tank reactors. It was assumed that both cascades are fed with the flux of raw materials with periodically changing the flow directions. Exemplary calculations were made upon the assumption that both fluxes flow counter-currently, and the flow reversal occurs at equal time intervals $\tau_p$. The analysis was primarily focused on the impact of time $\tau_p$ on the type of temperature and concentration oscillations in the system. In the course of numerical simulations the steady-state diagrams were designated, assuming that $\tau_p$ is a bifurcation parameter. The period of the appearance of windows of chaos in the diagram was found to be equal to the oscillation period of the state variables of the system without flow reversal. *N*-periodic oscillations of the system state variables without flow reversal may lead to the generation of *N*-periodic sequence of windows of chaos in the steady-state diagram with flow reversal.